\documentclass[%
superscriptaddress,
preprint,
showpacs,preprintnumbers,
 amsmath,amssymb,
aps,
]{revtex4-1}

\usepackage{graphicx}
\usepackage{color}
\usepackage{dcolumn}
\usepackage{bm}
\usepackage{hyperref}

\begin{document}

\title{Phosphorene and Transition Metal Dichalcogenide 2D Heterojunctions: Application in Excitonic Solar Cells}

\author{Vellayappan Dheivanayagam S/O Ganesan}
\affiliation{Engineering Science Programme, National University of Singapore, 9 Engineering Drive 1, Singapore 117575}

\author{Chun Zhang}
\affiliation{Department of Physics, National University of Singapore, Singapore 117542}

\author{Yuan Ping Feng}
\affiliation{Department of Physics, National University of Singapore, Singapore 117542}

\author{Lei Shen}
\email{shenlei@nus.edu.sg}
\affiliation{Engineering Science Programme, National University of Singapore, 9 Engineering Drive 1, Singapore 117575}

\date{\today}

\begin{abstract}
Using the first-principles GW-Bethe-Salpeter equation method, here we study the excited-state properties, including quasi-particle band structures and optical spectra, of phosphorene, a two-dimensional (2D) atomic layer of black phosphorus. The quasi-particle band gap of monolayer phosphorene is 2.15 eV and its optical gap is 1.6 eV, which is suitable for excitonic thin film solar cell applications. Next, this potential application is analysed by considering type-II heterostructures with single layered phosphorene and transition metal dichalcogenides (TMDs). These heterojunctions have a potential maximum power conversion efficiency of up to 12\%, which can be further enhanced to 20\% by strain engineering. Our results show that phosphorene is not only a promising new material for use in nanoscale electronics, but also in optoelectronics.

\end{abstract}

\maketitle


\section{INTRODUCTION}
The discovery of graphene in 2004 \cite{Novoselov2014Science} was a significant breakthrough in materials science and since then there has been sustained research interest in graphene and a rush to discover other stable two-dimensional (2D) materials. 2D materials have the thickness of one or a few atomic layers and have markedly different material properties than their bulk counterparts due to the quantum confinement effect. Since graphene, there have been several advances in the field of 2D materials such as the discovery of semiconducting monolayer transition metal dichalcogenides (TMDs) \cite{2_mak_lee_hone_shan_heinz_2010, 2_wang_kalantar-zadeh_kis_coleman_strano_2012}. The growing number and variety of 2D materials has fuelled interest in the use of 2D materials in the development of novel nanoscale devices.

A recent development is the experimental isolation of a single layer of bulk black phosphorus, also known as phosphorene \cite{Brent2014CC,Buscema2014NC,Castellanos-Gomez20142DM,Li2014NN,Liu2014AN,Qiao2014NC,Rodin2014PRL,Zhu2014PRL,Reich2014Nature,Churchill2014NN,Koenig2014APL,Fei2014NL,Fei2015PRB,Tran2014PRB,Tran2014PRB2,Guo2014JPCC,Peng2014PRB,Peng2014MRE,Wang2015PRB,Cai2014SR}. Like graphene, phosphorene was first obtained by mechanical exfoliation \cite{Li2014NN,Liu2014AN}. Liquid exfoliation, which is a scalable process, has also been demonstrated as a possible alternative means of producing phosphorene \cite{Brent2014CC}. Several experiments have demonstrated that phosphorene is a direct-gap semiconductor and also has a high hole mobility\cite{Li2014NN,Xia2014NC,Liu2014AN}. These characteristics make phosphorene attractive for use in electronic and optoelectronic devices \cite{Xia2014NC}. Field effect transistors (FET) based on few-layer phosphorene were shown to have high on/off ratios \cite{Li2014NN}. It also has the potential to be used as an anode material in lithium-ion batteries \cite{Li2015NL}. Another potential application is in thin film excitonic solar cells as phosphorene has a predicted band gap in the visible region \cite{Dai2014JPCL}. Excitonic solar cells (XSCs) based on some 2D materials, such as MoS$_2$, WSe$_2$, graphene, h-BN, SiC$_2$ and bilayer phosphorene, are potentially seen as the new generation of thin film solar cells\cite{1_tsai_su_chang_tsai_chen_wu_li_chen_he_2014, 2_pospischil_furchi_mueller_2014,Bernardi2012ACS,Dai2014JPCL,Zhou2013NL,Britnell2013Science}, and they might have higher efficiencies than existing XSCs which typically have less than 10\% efficiency \cite{1_green_emery_hishikawa_warta_dunlop_2015}. Till now, despite the limitations of fabrication methodologies for such 2D solar cells, there has been some progress on the fabrication of few-layer heterostructures such as graphene-WS${_2}$\cite{1_georgiou_jalil_belle_britnell_gorbachev_morozov_kim_gholinia_haigh_makarovsky_2012, 2_tan_avsar_balakrishnan_koon_taychatanapat_ofarrell_watanabe_taniguchi_eda_castro_neto_2014}, graphene-MoS${_2}$\cite{Britnell2012Science} and phosphorene-MoS${_2}$\cite{Deng2014ACS}.

In this paper we study the excited-state properties of monolayer phosphorene, and evaluate the viability of monolayer phosphorene as one building block of of an excitonic solar cell heterostructure. For the other building-block material in the heterostructure, the semiconducting monolayer TMDs, which have been extensively researched, are chosen. The semiconducting TMDs considered here include semiconducting MoS${_2}$, MoSe${_2}$, MoTe${_2}$, WS${_2}$, WSe${_2}$, WTe${_2}$, TiS${_2}$ and ZrS${_2}$. This paper is organized as follows: In \textbf{Sec. II}, we first introduce the structures of geometrically optimized monolayer phosphorene and computational details. \textbf{Sec. III} is the main part of results and discussions, which includes three subsections: In \textbf{SubSec. A}, we show the excited-state properties of monolayer phosphorene, including in quasi-particle band structures and optical spectra. In \textbf{SubSec. B}, we calculate the excited-state properties of 8 semiconducting TMDs. In \textbf{SubSec. C}, the band alignment of phosphorene and TMDs is presented. Then the power conversion efficiencies (PCE) of the phosphorene-TMD heterostructures are discussed. In \textbf{Sec. IV}, we conclude our studies.

\section{Structures and COMPUTATIONAL DETAILS}

\begin{figure}
\centering
\includegraphics[width=0.7\textwidth]{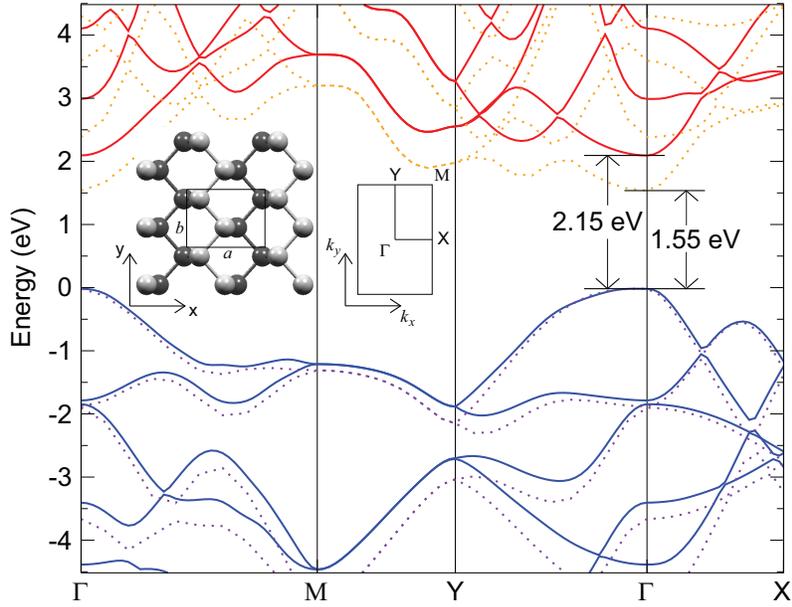}\\
\caption{(Color online) GW-calculated (solid lines) and HSE06-calculated (dash lines) band structure of monolayer phosphorene. The Fermi level is set on the top of the valance band. The ball and stick model of phosphorene (light color of the upper layer) with primitive cell and the reciprocal lattice with the high symmetry points are shown in inset.}
\label{fig1}
\end{figure}

Phosphorene has a puckered honeycomb structure as shown in inset of \textbf{Figure 1}. The underlying lattice is rectangular which leads to anisotropy in the band structure and optical properties. The calculated lattice constants are $a = 4.58$~\AA~and $b = 3.30$~\AA~which are consistent with results in literature \cite{Liu2014AN}. The calculations were performed using the Vienna Ab-initio Simulation Package (VASP) \cite{16_kresse_furth_1996, 15_kresse_furth_1996, 13_kresse_hafner_1993, 14_kresse_hafner_1994}, and a projector augmented wave (PAW) basis set was used \cite{1_blochl_1994, 2_kresse_joubert_1999}. Geometry optimisation was done using density function theory (DFT) with the generalised gradient approximation (GGA) using the Perdew-Burke-Ernzerhof (PBE) exchange correlation functional. A $14\times10\times1$ $k$-point grid was used for phosphorene while an $11\times11\times1$ $k$-point grid was used for TMDs. The geometry was relaxed until the force acting on the atoms was less than 0.01 eV/atom. To ensure that the interlayer interaction is negligible, the out of plane lattice parameter, which is perpendicular to the plane of the material, was set as at least 15 \AA.
Band gaps and band structures were calculated using the GW method. The band gaps/structures were also calculated through the screened exchange hybrid density functional by Heyd, Scuseria, and Ernzerhof (HSE06) for reference. A high number of empty conductions bands is necessary for the convergence of the absolute band positions\cite{Liang2013APL}. A total of 1024 bands were used for phosphorene and 1536 bands were used for TMDs. Single shot GW calculations, $G_{0}W_{0}$, were performed on phosphorene to obtain the ground state band energies. For TMDs, one eigenvalue update is performed to obtain the expected direct band gap of trigonal prismatic TMDs \cite{Shi2013PRB,Cheiwchanchamnangij2012PRB, Ramasubramaniam2012PRB, Qiu2013PRL}. The band structure was then interpolated from Wannier functions rather than evaluated directly at discrete $k$-points. This was done using the WANNIER90 library \cite{12_mostofi_yates_lee_souza_vanderbilt_marzari_2008} and the VASP2WANNIER interface. The optical gap was calculated by solving the Bethe-Salpeter equation (BSE). The GW method has more computational demands and the band gaps converge with a much smaller number of bands\cite{Liang2013APL}. Thus to streamline optical calculations, the $G_{0}W_{0}$ calculation was run with 128 bands for phosphorene and 192 bands for TMDs before solving the BSE. This produced the frequency dependent dielectric tensor that was used to calculate the absorption spectrum. By solving the BSE, electron-hole interactions such as excitons are accounted for in the dielectric tensor. For phosphorene the $x$- and $y$-components of the dielectric tensor were treated separately because of the anisotropy but for TMDs, the average value of the $x$- and $y$-components was used due to crystal symmetry.

\section{RESULTS AND DISCUSSIONS}
\subsection{Excited-state properties of phosphorene}

\begin{figure}
\centering
\includegraphics[width=0.75\textwidth]{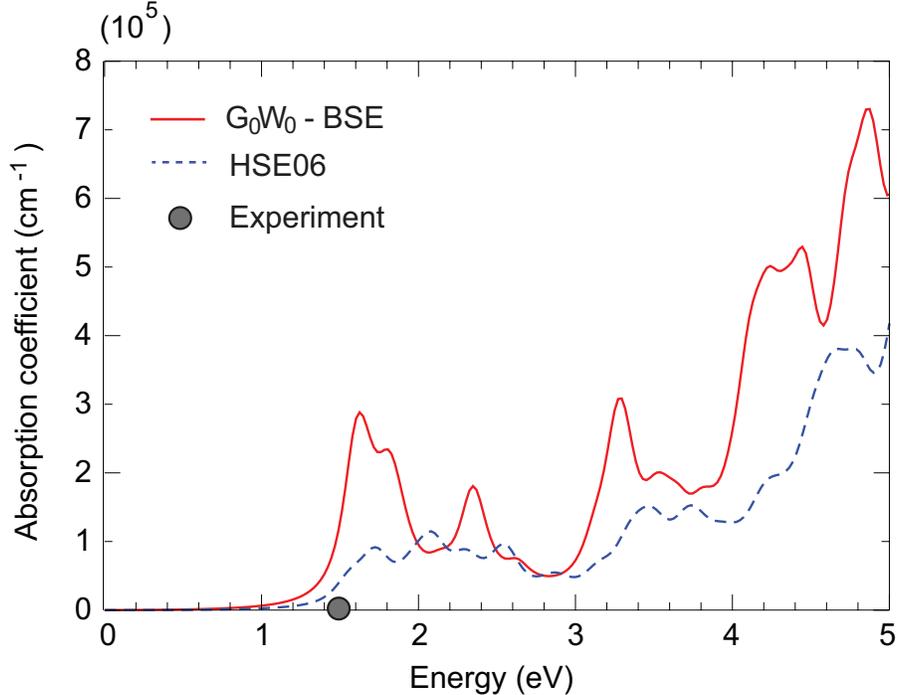}\\
\caption{(Color online) GW-calculated (solid lines) and HSE06-calculated (dashed lines) absorption spectrum of phosphorene for \emph{armchair} polarised light. The energy values of first absorption peak, observed in the experiment\cite{Li2014NN}, is presented by a black solid circle for reference.}
\label{fig2}
\end{figure}

\begin{figure}
\centering
\includegraphics[width=0.6\textwidth]{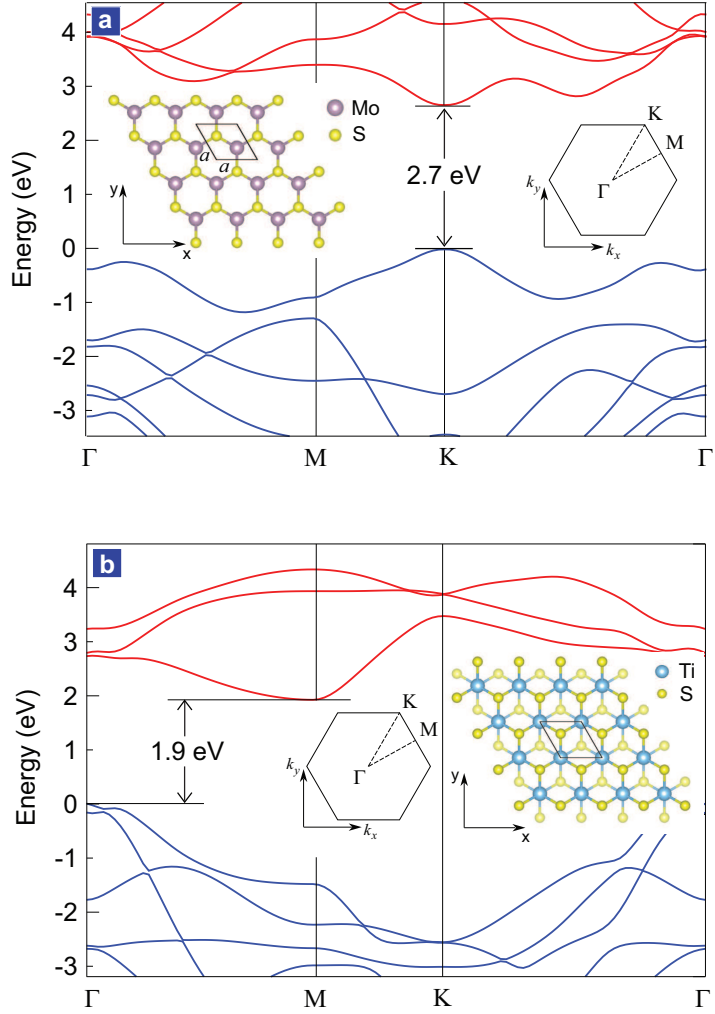}\\
\caption{(Color online) (a) The band structures of MoS${_2}$ with a ball and stick model of the trigonal prismatic geometry. (b) The band structures of TiS${_2}$ with a ball and stick model of the octahedral geometry. The reciprocal lattice with high symmetry points is shown in the inset. The Fermi level is set on the top of the valance band.}
\label{fig3}
\end{figure}
The calculated band structure of phosphorene is shown in \textbf{Figure 1}. The GW calculation predicts a band gap value of 2.15 eV and the gap is approximately located at the $\Gamma$ point, which is much higher than the DFT-calculated band gap of 0.91 eV \cite{Qiao2014NC}. The conduction band minimum (CBM) and the valence band maximum (VBM) are not exactly aligned in the GW-calculated band structure, but they are sufficiently close to be considered as a direct band gap \cite{Tran2014PRB2}. A similar computational approach by Tran \textit{et al}. predicted a band gap of 2.0 eV and a comparable profile of the band structure \cite{Tran2014PRB2}. The CBM position is around -4.25 eV with respect the the vacuum level. Besides the GW method, the hybrid density functional HSE06 was used to calculate the band structures and band gap for reference in \textbf{Fig. 1}. As can be seen, the HSE06-calculated band gap is 0.6 eV lower than the GW-calculated gap, which is in agreement with previous calculations \cite{Qiao2014NC}.

The optical gap of phosphorene is calculated using the GW-BSE approach and is determined to be 1.6 eV. This is the first optical peak of the absorption spectrum for light polarised along the armchair direction as seen in \textbf{Fig. 2}. This optical gap is slightly larger than the experimental photoluminescence measurement value of 1.45 eV of phosphorene \cite{Liu2014AN}, but smaller than the HSE06-calculated value of 1.8 eV. The optical gap of 1.6 eV is much lower than the electronic band gap of 2.15 eV which suggests that significant excitonic effects are present in phosphorene. The exciton binding energy of 0.55 eV in monolayer phosphorene is quite huge. Both the self-energy corrected large electronic gap and small optical gap of phosphorene indicate significant many-electron effect in phosphorene.

\subsection{Excited-state properties of TMDs}

Using a similar approach to the above, the geometry, band positions and band structures of 8 semiconducting TMDs are calculated. A summary of the lattice constants, absolute positions of the valence band maximum (VBM) and conduction band minimum for the different materials is shown in \textbf{Table 1}. The TMDs are categorised into either trigonal prismatic or octahedral TMDs, both of which have a hexagonal lattice but different coordination of the atoms within the unit cell. The two different structures are shown in the inset of \textbf{Fig. 3}. For trigonal prismatic TMDs, the lattice constant is largely determined by the size chalcogen atom which increases with atomic number (sulphur to selenium to tellurium). Furthermore, as can be seen, phosphorene has a large exciton binding energy compared to TMDs because of its unique quasi 1D band dispersions.

\begin{table*}
\centering
\caption{Summary of the lattice constants $a$ and $b$, the band gap $E_g$, valence band maximum $E_{VBM}$, conduction band minimum $E_{CBM}$ and optical gap $E_{opt}$}\label{tab:res}
\begin{tabular}{c c c c c c c c c}
\toprule
Material & Lattice & Structure & $a$ (\AA) & $b$ (\AA) & $E_g$ (eV) & $E_{VBM}$ (eV) & $E_{CBM}$ (eV) & $E_{opt}$ (eV) \\
Phosphorene & Rectangular & - & 4.58 & 3.30 & 2.15 & -6.40 & -4.25 & 1.6\\
MoS${_2}$  & Hexagonal & Trigonal Prismatic & 3.16 & - & 2.68 & -6.57 & -3.89 & 2.3\\
MoSe${_2}$ & Hexagonal & Trigonal Prismatic & 3.29 & - & 2.39 & -5.97 & -3.58 & 2.2\\
MoTe${_2}$ & Hexagonal & Trigonal Prismatic & 3.52 & - & 1.74 & -5.43 & -3.69 & 1.7\\
WS${_2}$   & Hexagonal & Trigonal Prismatic & 3.16 & - & 2.94 & -6.47 & -3.53 & 2.5\\
WSe${_2}$  & Hexagonal & Trigonal Prismatic & 3.26 & - & 2.70 & -5.84 & -3.14 & 2.6\\
WTe${_2}$  & Hexagonal & Trigonal Prismatic & 3.52 & - & 1.98  & -5.35 & -3.37 & 1.9\\
TiS${_2}$  & Hexagonal & Octahedral & 3.37 & - & 1.88 & -6.80 & -4.88 & -\\
ZrS${_2}$  & Hexagonal & Octahedral & 3.58 & - & 2.65 & -7.56 & -4.81 & -\\
\end{tabular}
\end{table*}

The trigonal prismatic TMDs have a direct band gap the the $K$ point [see \textbf{Supplementary Materials}]. They have the similar band structures. The band structure of MoS${_2}$ is shown in \textbf{Fig. 3(a)} as an example. For trigonal prismatic TMDs, the band gap decreases with increasing chalcogen atomic number. The CBM position also decreases with increasing chalcogen atomic number. These trends and the band gap values are consistent with other similar studies of these materials \cite{Gong2013APL,Liang2013APL,Kang2013APL, Ding2011PBCM,Shi2013PRB}. Octahedral TMDs have an indirect band gap between the $\Gamma$ and $M$ points [see \textbf{Supplementary Materials}]. This is also in agreement with band profiles calculated in previous studies. \cite{Liang2013APL,Gong2013APL,6_ivanovskaya_2005,Li2014RSC,Kang2013APL,Jiang2012JPCC}. The band structure of TiS${_2}$ is shown in \textbf{Figure 3(b)}.

\subsection{Band offset and PCE of excitonic solar cells}

\begin{figure}[htp]
\centering
\includegraphics[width=0.8\textwidth]{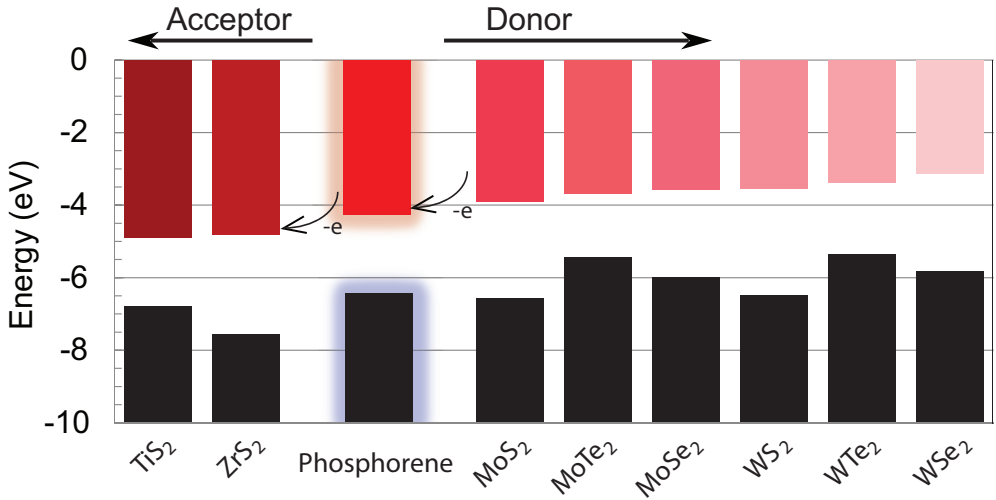}\\
\caption{(Color online) Band alignment of phosphorene with semiconducting monolayer TMDs.}
\label{fig6}
\end{figure}

Note that many important properties and potential device applications of semiconductors are not determined entirely by the band gap only. The band alignment and corresponding band offsets (the relative band-edge energies) of two or more semiconductors are other fundamental/critical parameters in the design of heterojunction devices \cite{Wei1998APL,Zhang2000PRL,Jiang2012JPCC,Kang2013APL,Liang2013APL}, for example, the 2D heterostructure devices for photocatalytic water splitting \cite{Jiang2012JPCC,Kang2013APL,Liang2013APL}, field effect transistors \cite{Gong2013APL} and \textit{p-n} diodes\cite{Deng2014ACS}. Chemical trends of the band offesets provide a useful tool for predicting catalytic ability of TMDs-based heterojunctions. \textbf{Figure 4} shows the band alignment (using the vacuum level as reference) of phosphorene with 8 semiconducting monolayer TMDs. It can be seen that the CBM of trigonal prismatic TMDs is higher than that of monolayer phosphorene, while the CBM of octahedral TMDs is lower than phosphorene. Thus, trigonal prismatic TMDs could function as the donor whereas octahedral TMDs would function as the acceptor in heterostructure with phosphorene. The optical gap of 8 TMDs is calculated, which is determined from the absorption spectrum. In most cases, the optical gap is slightly lower than the electronic band gap as shown in \textbf{Table 1}. Notice that because of strong many-electron effects in some 2D materials, such as phosphorene and MoS$_2$, we might not get the accurate band offset parameters in some cases without considering the excited-state effect. For example, the HSE-calculated CBM band energy of monolayer MoS$_2$ is -4.21~eV \cite{Guo2014JPCC} (-4.25~eV \cite{Kang2013APL}) and phosphorene is -3.94~eV \cite{Guo2014JPCC} (-3.92~eV \cite{Cai2014SR}). Thus, phoshporene (MoS$_2$) is the donor (acceptor) in the phosphorene-MoS$_2$ 2D heterojunction \cite{Guo2014JPCC} and the predicted PCE is 17.5\% \cite{Guo2014JPCC}. However, if we consider many-electron effects, the GW-calculated CBM band energy of monolayer MoS$_2$ is -3.89~eV (-3.74~eV \cite{Liang2013APL}) and phosphorene is -4.25~eV. Interestingly, phoshporene (MoS$_2$) becomes the acceptor (donor) instead and the predicted PCE is reducing to 10\%. Regarding to other 2D materials without strong many-electron effects, both HSE and GW can give similar band offset \cite{Liang2013APL,Kang2013APL}.

A model developed by Scharber et al. for organic solar cells \cite{1_scharber_2006} and later adapted for exciton based 2D solar cells \cite{Bernardi2012ACS,Dai2014JPCL} is used to predict the maximum PCE, $\eta$, based on the fill-factor, $\beta_{FF}$, open circuit voltage, $V_{oc}$, and short circuit current, $J_{sc}$.
\begin{align}
\eta = \frac{\beta_{FF}V_{oc}J_{sc}}{P_{\mathrm{solar}}}
\end{align}
where $P_{\mathrm{solar}}$ is the total incident solar power per unit area based on the Air Mass (AM) 1.5 solar spectrum \cite{1_astm_2012, 2_RReDC_2014}. The fill factor is the ratio of power output at the maximum power point to the product of the open circuit voltage and the short circuit current. The fill factor is estimated to be $0.65$ from literature. The $V_{oc}$, in units of V, and $J_{sc}$, in units of A/$\mathrm{m}^2$, are estimated in the limit of 100\% external quantum efficiency as
\begin{align}
V_{oc} &= \frac{1}{e}\left( E_{opt}^d - \Delta E_{CBM} - 0.3 \right) \label{eq:voc}\\
J_{sc} &= e\int_{E_{opt}^d}^\infty \frac{(\hbar \omega)}{\hbar \omega} d \hbar \omega
\end{align}
where $e$ is the elementary charge, $E_{opt}^d$ is the donor optical gap, $\Delta E_{CBM}$ is the conduction band offset and $P(\hbar \omega)$ is AM 1.5 solar spectrum. In equation \ref{eq:voc}, the constant 0.3 eV is an empirical parameter that estimates losses due to energy conversion kinetics.

\begin{figure}[htp]
\centering
\includegraphics[width=0.7\textwidth]{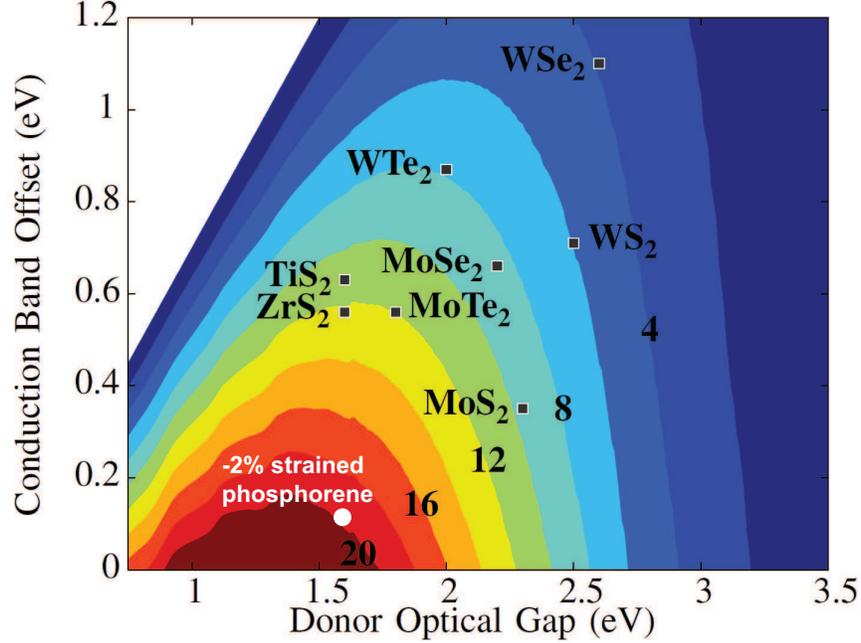}\\
\caption{(Color online) Power conversion efficiency of potential excitonic thin film solar cell heterojunctions. The labels indicate the material that complements phosphorene.}
\label{fig7}
\end{figure}

Using this model the TMDs are paired with phosphorene. The material with the lower CBM is the acceptor and the material with the higher CBM is the donor. Phosphorene is the donor when paired with both octahedral TMDs and the acceptor when paired with trigonal prismatic TMDs. The maximum PCE values for these eight heterostructures are marked on \textbf{Fig. 5}. Of the eight heterostructures, phosphorene-ZrS${_2}$ and MoTe${_2}$-phosphorene have the highest PCE value of 12\%. This efficiency is higher than that achieved by existing excitonic solar cells, and comparable to the proposed 2D g-SiC$_2$/GaN (14.2\%), PCBM/CBN (10-20\%)\cite{Bernardi2012ACS}, and bilayer-phosphorene/MoS$_2$ (16-18\%) solar cells.

\begin{figure} [htp]
\centering
\includegraphics[width=0.8\textwidth]{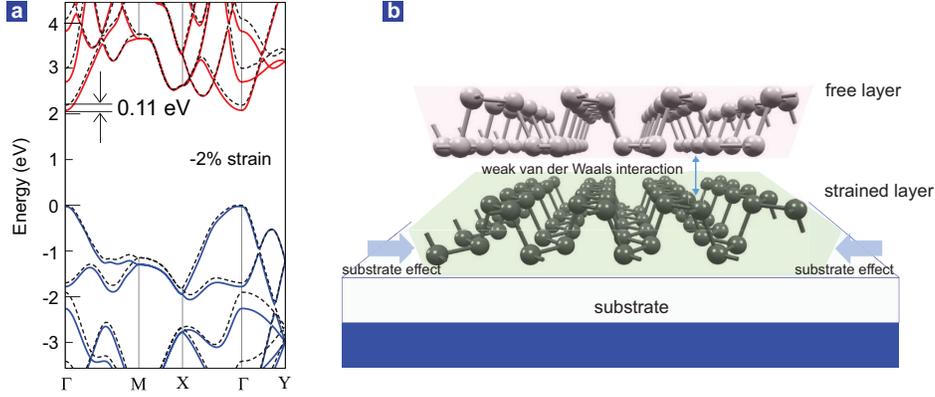}\\
\caption{(Color online) (a) The band structure of phosphorene under applied 2\% armchair compressed strain. The Fermi level is set on the top of the valance band. The dashed black lines are the band structure of phosphorene without strain for reference. (b) The schematic diagram of bilayer strained phosphorene-free phophorene heterostructure. The strain of the 1$^{\textrm{st}}$ layer can be induced by the substrate, whereas the 2$^{\textrm{nd}}$ layer is unstrained because of the weak van der Waals interaction.}
\label{fig7}
\end{figure}
Actually, a further observation based on this model is that a solar cell with a phosphorene donor could have maximum PCE values of up to about 20\% with an appropriate choice for the acceptor. The conduction band offset (CBO) between phosphorene and TiS${_2}$ is 0.63 eV while the CBO between phosphorene and ZrS${_2}$ is 0.56 eV for the two cases here where monolayer phosphorene is the donor. This translates to a significant drop in $V_{oc}$ in the model which results in a lower PCE. In addition to looking out for new materials that have a better band alignment with phosphorene, means of tuning the properties of both the donor and acceptor can be considered. For example, strained phosphorene may be used as the acceptor material because the strain effect is a well-known method to tune the band structure of materials \cite{Fei2014NL,Fei2015PRB}. The band structures of different strained monolayer phosphorene are shown in \textbf{Supplementary Materials}. Here, we choose the phosphorene with 2\% compressed strain (along the armchair direction) as a donor. Its band structure is shown in \textbf{Fig. 6}. For comparison, the band structure of phosphorene without strain is also presented in the same figure. As can be seen, the position of CMB can be effectively tuned by the strain effect. 2\% compressed armchair strain can shift the CBM down to the Fermi level around 0.11 eV because the compressed strain enhances the interaction of hybridized $s-p$ orbitals of P atoms, which contribute to the CBM. The calculated PCE value of -2\% strained-phosphorene/phosphorene can be around 20\% as shown by the white solid circle in \textbf{Fig. 5}. Based on the theoretical studies of mechanical properties of monolayer phosphorene, the mechanical stability of phosphorene can be up to under 30\% strain \cite{Wei2014APL,Peng2014PRB}. The 2\% compressed strain of a monolayer phorphorene can be realized by the substrate effect in the experiment [see Fig. 6(b)]. Meanwhile, there is no strain on the second deposited monolayer phosphorene because of the weak van der Waals interaction between the two phosphorene layers. This can realize the strain/non-strain phosphorene heterostructure.

Alternatively multilayer structures of phosphorene or TMDs may also be considered to increase the overall power conversion per unit area. Our calculated optical gap of phosphorene is 1.6 eV which is at the edge of the infrared region. Therefore, heterostructures with a phosphorene donor may be considered for multijunction solar cells absorbing photons across the entire visible spectrum. Such cells incorporate several junctions that aim to absorb different portions of the solar spectrum so as to maximise total absorption [See \textbf{Supplementary Materials}]. Given the anisotropy of phosphorene, the stacking orientation in multilayer structures may be a significant factor \cite{Dai2014JPCL}.

\section{CONCLUSIONS}
In conclusion, through GW calculations, the band structures and optical spectra of monolayer phosphorene have been calculated. The electronic gap of 2.15 eV and the optical gap of 1.6 eV are desirable for solar cell applications because of the strong exciton binding energy. When paired with ZrS${_2}$ or MoTe${_2}$ the power conversion efficiency of the excitonic solar cells can be as high as 12\%. There is further potential to improve the PCE of phosphorene based solar cells substantially by tuning the materials, such as through the strain and multi-stacking effect, to achieve a better band alignment. 

\section{ACKNOWLEDGEMENT}

Authors thank Minggang Zeng for his helpful discussion. This project was supported by the Computational Condensed Matter Physics Laboratory, NUS. Computational resources were provided by Centre for Advanced 2D Materials, NUS.

%

\end{document}